\documentclass[manuscript]{emulateapj}
\usepackage{amsmath}
\usepackage{graphics}      
\usepackage{graphicx}

\begin{document}
\newcommand{\kms}{\mbox{km~s$^{-1}$}}
\newcommand{\s}{\mbox{$''$}}
\newcommand{\mloss}{\mbox{$\dot{M}$}}
\newcommand{\mdot}{$\dot{M}  $}
\newcommand{\my}{\mbox{$M_{\odot}$~yr$^{-1}$}}
\newcommand{\ls}{\mbox{$L_{\odot}$}}
\newcommand{\ms}{\mbox{$M_{\odot}$}}
\newcommand{\rs}{\mbox{$R_{\odot}$}}
\newcommand{\msun}{\mbox{$M_{\odot}$}}
\newcommand{\rsun}{\mbox{$R_{\odot}$}}
\newcommand{\lsun}{\mbox{$L_{\odot}$}}
\newcommand{\hst}{{\it HST}}
\newcommand{\gsim}{\raisebox{-.4ex}{$\stackrel{>}{\scriptstyle \sim}$}}
\newcommand{\lsim}{\raisebox{-.4ex}{$\stackrel{<}{\scriptstyle \sim}$}}
\newcommand{\teff}{\mbox{$T_{eff}$}}
\newcommand{\lya}{\mbox{Ly$\alpha$}}
\newcommand{\fluxu}{\mbox{ergs\,cm$^{-2}$\,s$^{-2}$}}
\newcommand{\keunit}{\mbox{ergs}}
\newcommand{\momunit}{\mbox{\ms\,\kms}}

\title{Binarity and Accretion in AGB Stars: HST/STIS Observations of UV Flickering in Y\,Gem}
\author{R. Sahai\altaffilmark{1},  C. S{\'a}nchez Contreras\altaffilmark{2}, A. Mangan\altaffilmark{1}, J. Sanz-Forcada\altaffilmark{2}, 
C. Muthumariappan\altaffilmark{3}, M. J. Claussen\altaffilmark{4}
}
\altaffiltext{1}{Jet Propulsion Laboratory, MS\,183-900, California Institute of Technology, Pasadena, CA 91109}
\altaffiltext{2}{Astrobiology Center (CSIC-INTA), ESAC campus, E-28691 Villanueva de la C\~anada, Madrid, Spain}
\altaffiltext{3}{Indian Institute of Astrophysics, Bangalore 560034, India}
\altaffiltext{4}{National Radio Astronomy Observatory, 1003 Lopezville Road, Socorro, NM 87801}


\begin{abstract}
Binarity is believed to dramatically affect the history and geometry
of mass loss in AGB and post-AGB stars, but observational evidence of binarity is sorely lacking. As part of a project to search 
for hot binary companions to cool AGB stars using the GALEX archive, we discovered a late-M star, Y\,Gem, to be a source of
strong and variable UV and X-ray emission. Here we report UV spectroscopic observations of Y\,Gem obtained with the {\it Hubble Space Telescope} that show 
strong flickering in the UV continuum on time-scales of $\lesssim20$\,s, characteristic of an active accretion disk. 
Several UV lines with P-Cygni-type profiles from species such as Si\,IV and C\,IV 
are also observed, with emission and absorption features that are red- and blue- shifted by velocities of 
$\sim$500\,\kms~from the systemic velocity. Our model for these (and previous) observations is that material from the primary star 
is gravitationally captured by a companion, producing a hot accretion disk. The latter powers a fast outflow 
that produces blue-shifted features due to absorption of UV continuum emited by the disk, whereas the red-shifted emission features arise in heated infalling material from the primary. The outflow velocities support a previous inference by Sahai et al. (2015) that Y\,Gem's companion is a low-mass main-sequence star. 
Blackbody fitting of the UV 
continuum implies an accretion luminosity of about 13\,\ls, and thus a 
mass-accretion rate $>5\times10^{-7}$\,\my; we infer that Roche lobe 
overflow is the most likely binary accretion mode for Y\,Gem.
\end{abstract}


\section{Introduction}
One of the biggest challenges for 21st century stellar astronomy is a comprehensive 
understanding of the impact of binary interactions on stellar evolution. Close binary interactions are expected to dominate a substantial
fraction of stellar phenomenology -- e.g., cataclysmic variables, type
Ia supernovae progenitors, and low and high-mass X-ray binaries.

Binary star interactions are specifically believed to underlie the formation of the overwhelming majority of Planetary
Nebulae (PNe), which represent the bright end-stage of most stars in the Universe. Such interactions are likely the key
to the resolution of the long-standing puzzle: although PNe (and their progenitors, pre-PNe [PPNe]) evolve from
slowly-expanding ($V_{exp}\sim$\,5--15\,\kms) spherically-symmetric circumstellar envelopes (CSEs) of AGB stars, modern
surveys reveal that the vast majority of the former deviate strongly from spherical symmetry, showing a dazzling variety
of elliptical, bipolar and multipolar morphologies and fast, collimated outflows ($V_{exp}\gsim$\,50--100\,\kms) (e.g., Sahai \& Trauger 1998; Balick \& Frank 2002; Sahai et al. 2011a).

A close association between jets and binary interaction involving a red giant or AGB primary is exemplified by symbiotic stars, a small class of objects in which the optical spectrum shows features of TiO (showing the presence of a cool red
giant primary), but also optical emission lines e.g., H\,I, He\,II, and [OIII]. In such objects, a compact star, usually a white-dwarf (WD), 
accretes matter from the giant primary. Prime examples of symbiotic stars with jets are R\,Aqr and CH\,Cyg (e.g., Corradi et al. 1999).

However, there is a lack of observational evidence of widespread binarity in AGB stars. We have therefore 
been using UV and X-ray observations as new probes of accretion-related phenomena in AGB stars -- archival surveys in the UV using the GALEX database 
show that a large fraction of AGB stars show FUV emission, with relatively high FUV/NUV flux ratios and significant variability, likely resulting from 
variable accretion associated with a companion (Sahai et al. 2008,\,2011b,\,2015,\,2016). Roughly, about 40\% of objects surveyed show 
strong, variable, X-ray emission that likely arises in an accretion disk around a compact companion (see Sahai et al. 2015: Setal15).
In this paper, we report stochastic variations on $\lsim20$\,s time-scales in the UV spectrum of our best-studied UV/X-ray emitting star, Y\,Gem, 
similar to the flickering phenomenon seen in other well known classes of accreting binaries. These variations  
provide independent and robust evidence of binarity in this object.

Until recently, Y\,Gem lacked a significant measurement of the parallax (its Hipparcos parallax/error is $1.30$\,mas/$1.38$\,mas: van Leeuwen 2007), 
and we have adopted, in our previous studies of this object, a distance of $D=0.58$\,kpc, as inferred from its K-band magnitude, 
following Kahane \& Jura (1994) who assume that late-M semi-regular stars have absolute magnitudes $M_K = -7.6$  (Sahai et al. 2011b: Setal11). In the recent 
GAIA Data Release 2 (Lindegren et al. 2018, {\it in prep}), its measured parallax is $1.4997\pm0.1561$\,mas, giving a distance of $0.67\pm0.07$\,kpc. Since this is roughly within 1$\sigma$ of our 
adopted value, we have conservatively kept our original distance estimate for this paper, especially since the DR2 results 
have yet to be subjected to scrutiny by the astrophysical community. The slightly larger distance, if correct, would imply a $\sim$30\,\% increase in estimated luminosities, and does not 
affect our conclusions.

The plan of the paper is as follows. We describe the observational setup in \S\,\ref{obs}, and the data reduction and calibration procedures in \S\,\ref{reduce}. The observational results and 
their analysis are described in \S\,\ref{anal}, with focus on the spectral time-variability (\S\,\ref{timevar}), the UV continuum and its modeling (\S\,\ref{contmod}), and 
the line profiles (\S\,\ref{line_prof}). We present a new geometrical model to explain 
our results, and discuss their implications for binarity and accretion in Y\,Gem in \S\,\ref{discus}. Our 
main conclusions are summarized in \S\,\ref{conclude}.

\section{Observations}\label{obs} We obtained UV spectra of Y\,Gem with the Space Telescope Imaging Spectrograph (STIS) on board the {\it Hubble Space Telescope} ($HST$) between UT 22:05 10-11-2016 and 00:47 10-12-2016, using the G140L and G230L gratings, and 
the $52{''}\times0.2{''}$ aperture in {\it Time-Tag} mode. The slit width fully encompasses the entire source, which is expected to be unresolved (Setal15). The observing sequence consisted of two pairs of G140L exposures, followed by two pairs of G2300L exposures; this sequence was repeated a second time, providing 4 separate exposures each in 
the two wavelength bands ($\sim$1120\,\AA--1700\,\AA~and $\sim$1600\,\AA--3150\,\AA). An observation log is given in Table\,\ref{logobs}. We obtained similar data during a following epoch (April 2017), but 
during these observations the source was much brighter, causing the time-tag buffer to overflow, so the absolute flux levels are not reliable. These data 
will be discussed in a subsequent paper.

\section{Data Reduction}\label{reduce} 
In order to observe changes in the spectra over time intervals smaller than the full integration time of each exposure, we analysed our data in {\it Time-Tag} mode, 
following the procedure described in the STIS Time-Tag Analysis Guide (Dashevsky, Sahu, \& Smith 2000).  We used IRAF to run the ``{\it odelaytime}" command, which applied Heliocentric and Barycentric time corrections, followed by 
``{\it inttag}" which created a raw datafile, with multiple sub-exposures of a chosen duration. Each sub-exposure used data only in the Good Time Intervals (GTI).  The subexposures at the end of each GTI had shorter durations resulting in relatively larger errors in the associated fluxes. In our analysis below, we chose a sub-exposure time of 20\,s in order to obtain a good signal-to-noise ratio while still maintaining a large number of subexposures for our timing analysis. We ran {\it calstis} in {\it PyRAF} to create FITS-format files of the calibrated subexposure data.

\section{Results and Analysis}\label{anal}
\begin{figure*}[hbt]
\resizebox{1.0\textwidth}{!}{\includegraphics{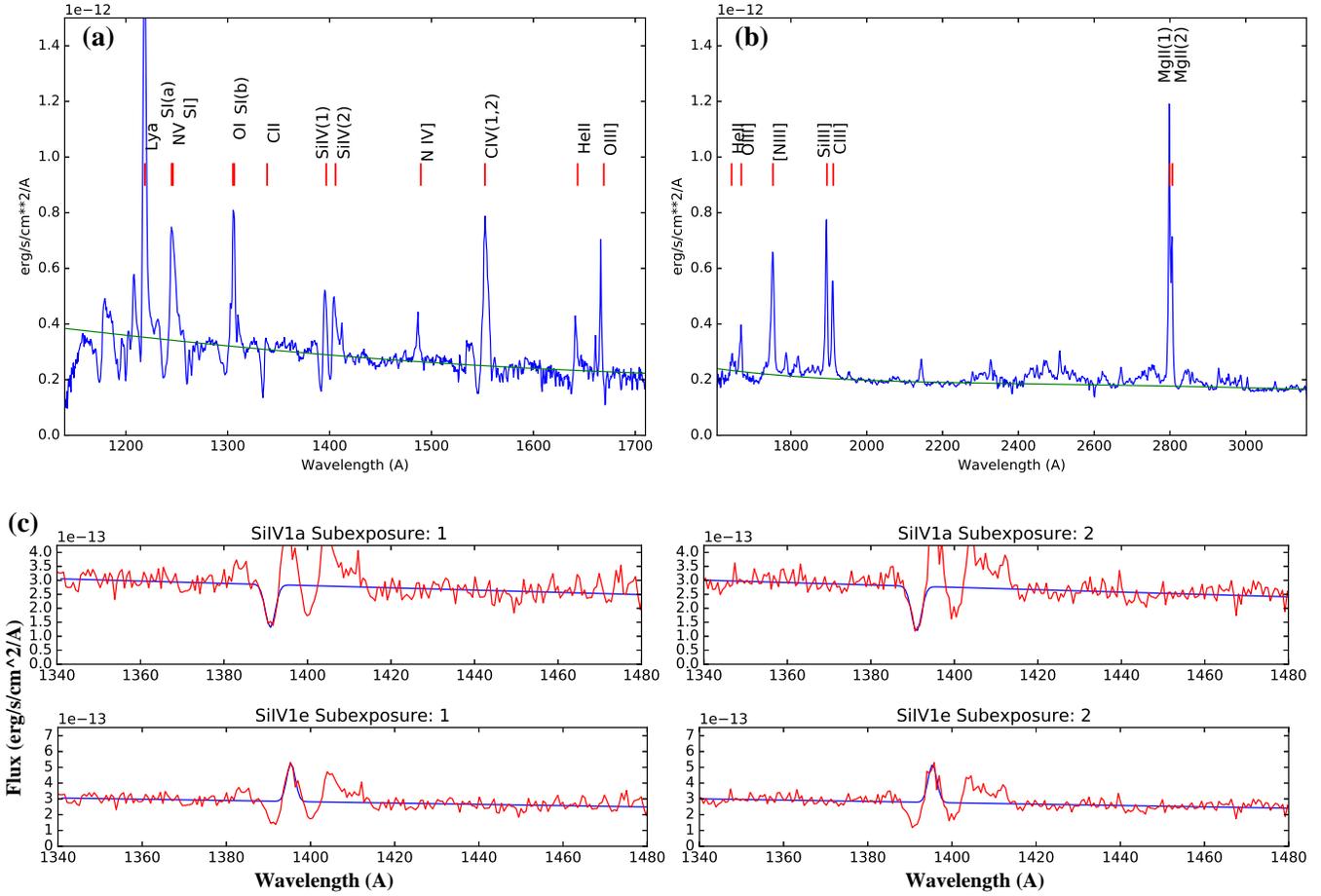}}
\caption{Representative STIS/UV spectra of Y\,Gem observed using gratings (a) G140L (exposure 20) and (b) G230L  (exposure 30). Exposures 20 and 30 were used to calculate the parameters for the spectral lines in Table\,\ref{lines}. The green curves show a model 
fit consisting of two blackbody components, characterised by \teff$ = 35,500$\,K, $L = 6.3$\,\ls, and \teff$ = 9,400$\,K, $L = 6.7$\,\ls. (c) 
STIS/UV spectra of Y\,Gem in the vicinity of the Si\,IV(1) line for the first two 20\,s subexposure in exposure 20. The blue curve shows Gaussian line-profile fits (together with a linear baseline) to the 
absorption (top) and emission (bottom) features. Such fitting has been used to calculate the parameters for the spectral lines in Table\,\ref{lines}.
}
\label{ygemspec20}
\end{figure*}

The G140L and G230L spectra show the presence of continuum emission and many strong emission lines, some with blue-shifted absorption features resulting in P-Cygni-type profiles (e.g., Castor 1970), from ionized levels of abundant metals such as 
C, O, N, Si, and Mg.  No molecular H$_2$ lines are seen. Representative spectra and line-lists are given in Fig.\,\ref{ygemspec20}a,b and Table\,\ref{lines}, respectively. For both the sub-exposure and full exposure spectra, 
we used small line-free wavelength windows around each strong line feature to subtract an underlying linear continuum, followed by Gaussian line fitting to determine the line center, Full Width at Half Maximum (FWHM) and flux of emission and absorption features (e.g., Fig.\,\ref{ygemspec20}c). 
The instrumental resolution, which varies from 1.7 (2.2) to 1.4 (2.1)\,\AA~over the wavelength range of the G140L (G230L) grating, has been 
deconvolved from the line-width values in Table\,\ref{lines}.  Representative line fitting results are given for exposures 20 and 30 in Table\,\ref{lines}.

\subsection{Time Variations}\label{timevar}
\begin{figure*}[hbt]
\resizebox{1.0\textwidth}{!}{\includegraphics{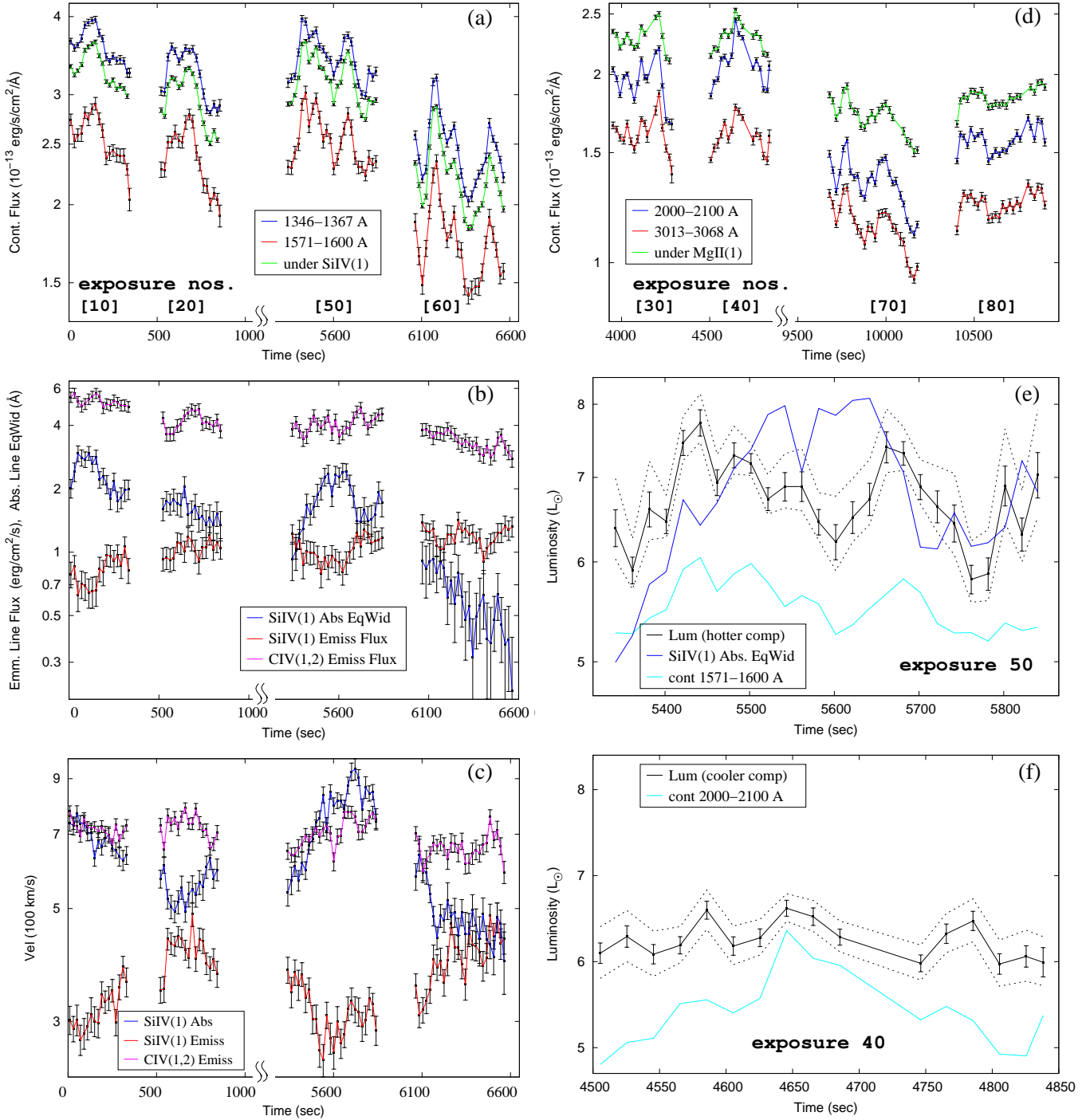}}
\caption{Short-term time variations in the continuum and lines observed in Y\,Gem: (a) line-free continuum in the bands $1346-1367$\,\AA~(blue) and $1571-1600$\,\AA~(red), and continuum underlying the Si\,IV(1) line (green), (b) the equivalent-width of 
the Si\,IV(1) absorption line (blue) and the flux of the Si\,IV(1) (red) and C\,IV(1,2) emission lines (pink). The emission-line fluxes have been scaled up by a factor $1.5\times10^{12}$, (c) the Doppler-shifts (absolute values) of the lines in panel $b$: the Si\,IV(1) absorption feature is blue-shifted, whereas the Si\,IV(1) and C\,IV(1,2) emission features are red-shifted, (d) the line-free continuum in the bands $2000-2100$\,\AA~(blue), $3013-3068$\,\AA~(red) and continuum underlying the Mg\,II(1) line (green; scaled up by factor 1.1 for clarity). Error bars in panels $a-d$ are $\pm1\sigma$. The light curves in panels $a-c$  and $d$ are, respectively, extracted from the G140L exposures 10,20,50,60 and G230L exposures 30,40,70,80. (e) the hotter blackbody's luminosity (black), derived using a two-blackbody fit to the continuum observed in the subexposures within exposure 50: the dashed curves show upper and lower bounds on the luminosity due to the estimated uncertainties in the cooler blackbody's luminosity and temperature. The 
$1571-1600$\,\AA~continuum, scaled up by a factor $1.1\times10^{7}$ (cyan), and the square-root of the  
Si\,IV(1) absorption line equivalent width, scaled up by a factor $5.2$ (blue), are shown for comparison, (f) as in panel $e$, but for the cooler blackbody's luminosity, derived from fitting the subexposures within exposure 40 (the $2000-2100$\,\AA~continuum, scaled up by a factor $2.6\times10^{13}$ (cyan), is shown for comparison).
}
\label{contlumtemp}
\end{figure*}

Our analysis of the sub-exposure spectra reveal fast variations in the continuum UV flux of Y Gem at all wavelengths (Fig.\,\ref{contlumtemp}a,d). 
The continuum light-curves, extracted from line-free regions, and that underlying the different spectral lines, are quite similar, with significant variations seen across successive sub-exposures, implying variability on timescales of 20\,s or less. The maximum peak-to-peak amplitude 
of the continuum variations is 0.53 mag (exposure 60). 
Using the ratio between observed and expected root-mean-square variability ($s$ and $s_{exp}$, respectively) 
to quantify the significance of stochastic variations in the light curves (see e.g., Nu{\~n}ez et al. 2016), we find $s/s_{exp}\sim6-8$ for several line-free continum windows in the G140L and G230L spectra. Periodogram analysis reveals no specific period; the variations appear to be stochastic in nature.

Fluctuations in the spectral-line parameters are also seen, but the uncertainties in the fitted parameters are larger and so in general, the variability is less significant than in the continuum. However, we do find significant fractional variations in the Si\,IV(1) absorption line equivalent-width (Fig.\,\ref{contlumtemp}b), and the Si\,IV(1) absorption and emission line Doppler-shifts (Fig.\,\ref{contlumtemp}c). In addition, the amplitude of the blue-shift for the Si\,IV(1) absorption feature appears to be anti-correlated with the red-shift of the Si\,IV(1) emission feature. 
In contrast, the Si\,IV(1) and C\,IV(1,2) emission line fluxes show much smaller variations within each exposure.

During the period covering exposures 10--60 ($\sim1.7$\,hr), excluding that covered by exposure 50, we find a general decline in the continuum fluxes, 
and the Si\,IV(1) absorption equivalent-width and blue-shift; however, these, together with the continuum fluxes, all deviate upward from the above trend during exposure 50. 
The most striking decline (by factors $>5$) is seen in the Si\,IV(1) equivalent-width, from the end of exposure 50 to the end of exposure 60. Correspondingly, 
the FUV continuum declines to its  lowest level in exposure 60. The Si\,IV(1) absorption line equivalent-width shows a broad peak at an epoch that lies between  
the two peak luminosity epochs in exposure 50. We interpret these correlations in the context of a geometric model in \S\,\ref{discus}.

We note that much smaller-amplitude photometric variations have been recently reported at  
optical wavelengths (e.g., 0.06 mag peak-to-peak in the $u^{\prime}$ band with a typical timescale of 10\,min) for Y\,Gem using 
ground-based differential photometry (Snaid et al. 2018). Although a comparison with differential photometry of two field stars indicates that these 
variations are significant, the effects of aperture photometry in variable seeing conditions can produce spurious fluctuations that are not easily quantified. 
For example, in the Snaid et al. data set, large spurious fluctuations are clearly noticeable in both the differential photometry and the seeing $\sim10-20$\,min into the observation time stream, but smaller 
seeing fluctuations are present during the rest of the time stream, and may be contributing to the photometric variations as well.

\subsection{Continuum Modeling}\label{contmod}
Single blackbody fits (using least-squares minimisation) to the G140L spectra from exposures 10,20,50, and 60 give values for the luminosities of $L\sim6.9, 6.0, 6.4$, and 4.5\,\ls~and effective temperatures of 
$\teff\sim31440, 29600, 29540$, and 29980\,K, respectively. However, these blackbodies  
do not provide sufficient flux at the longer wavelengths (i.e., $\gsim1800$\,\AA) covered in the G230L spectra; a significantly cooler blackbody is needed to fit the latter. For example, single blackbody fits to 
the G230L spectra from exposures 30,40,70, and 80 give values for the luminosities of $L\sim7.0, 7.0, 4.9$, and 5.5\,\ls~and effective temperatures of 
$\teff\sim13140, 13800, 13400$, and 13450\,K, respectively. Thus, whereas the effective temperatures do not change much with time, the luminosities derived for both the hotter 
and cooler blackbodies above show significant changes which are similar to those seen in the continuum level in the G140L and G230L exposures. We have not included dust extinction 
in our modeling because there is no measurable dust excess in Y Gem (Setal11).

In order to constrain the values of $L$ and $\teff$ for each of these blackbodies and their variations more accurately, we construct a model that combines two blackbodies to fit the G140L and G230L spectra.
Since the G140L and G230L spectra are 
not co-eval, and the fits above indicate that both $L$ and $\teff$ can vary with time, we use exposures 
40 (for G230L) and 50 (for G140L) as these were taken closest to each other in time. We find that the G140L spectrum requires $L(h) = 6.8$\,\ls~and $\teff(h)\sim36600$\,K, and a 
cooler blackbody with $L(c)\sim6.3$\,\ls, $\teff(c)\sim9940$\,K. 
Thus $L(h)$ and $\teff(h)$ respectively increase by about 6\% and 24\%, and $L(c)$ and $\teff(c)$ respectively decrease by about 29\% and 24\%, over the corresponding single blackbody fits.
The two-blackbody fits for the other pairs of successive G140L+G230L exposures (20+30 or 60+70) produce very similar 
values of $L(h)$, $\teff(h)$, $L(c)$, and $\teff(c)$, but these are much more uncertain as the G140L+G230L exposure pairs have very long time-gaps between them (about 1\,hr).  

We investigate luminosity and temperature changes in both blackbodies on the short variability timescales that we find for the continuum, as follows. 
First, we  determine  $L(h)$ and $\teff(h)$ as a function of subexposure in exposure 50 using two-blackbody fits, with $L(c)$ and $\teff(c)$ set to their 
(average) values for exposure 40 (as determined above). Then, in order to account for the variability of $L(c)$  and $\teff(c)$ during exposure 50, we follow the following procedure.  
First, using a single-blackbody fit to the subexposure spectra in exposure 40, we find that the peak-to-peak variations
in $L(c)$  and $\teff(c)$ are about $\pm10$\% and $\pm3$\%, respectively -- using these to set upper and lower bounds on the average value of 
$L(c)$ and $\teff(c)$, we determine the upper and lower bounds on the variability of
$L(h)$ and $\teff(h)$. We find significant variations in $L(h)$ (Fig.\,\ref{contlumtemp}e), but not in $\teff(h)$. A similar analysis for the cooler blackbody shows that 
the variations in $L(c)$ are marginally significant (Fig.\,\ref{contlumtemp}f), and no significant variations are found for $\teff(c)$. The variations in $L(h)$ and $L(c)$ follow the variations in the continuum fluxes (Figs.\,\ref{contlumtemp}e,f).

\subsection{Line Profiles}\label{line_prof}
\begin{figure*}[hbt]
\rotatebox{270}{\resizebox{0.5\textwidth}{!}{\includegraphics{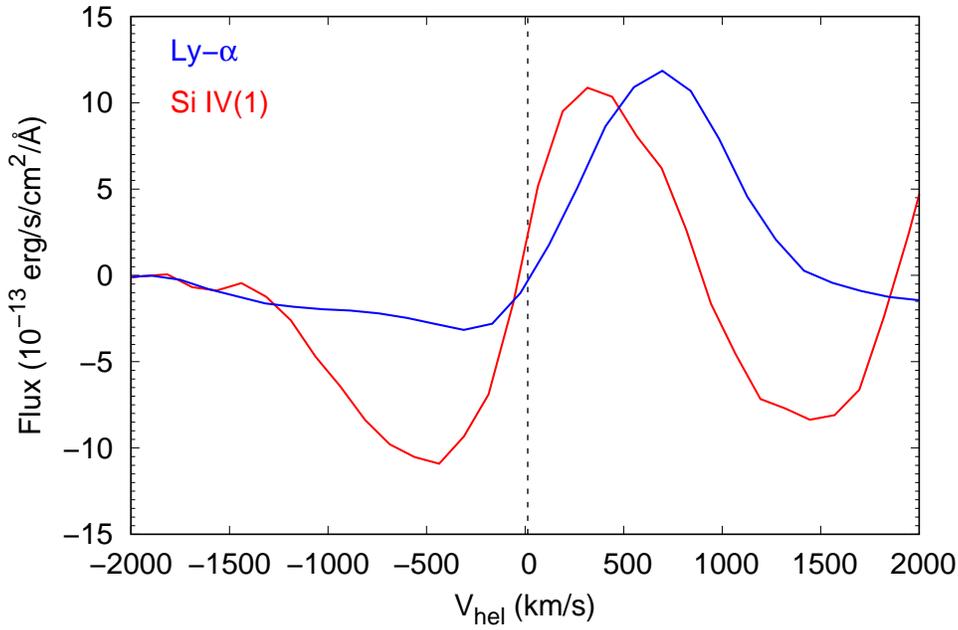}}}
\caption{Ly$\alpha$ ({\it blue}) and SI\,IV(1) ({\it red}) line profiles (continuum-subtracted) from exposure 20. Dashed vertical line shows the systemic velocity.
}
\label{siiv1_lya}
\end{figure*}

All emission lines are red-shifted relative to the systemic velocity ($V_{hel}=13.5$\,\kms); the velocities lie in the range $V_{hel} \sim 400-700$ ($150-250$) \,\kms~for lines 
observed with the G140L (G230L) grating (Table\,\ref{lines}, Fig.\,\ref{siiv1_lya}). Prominent P-Cygni-type absorption features are seen in lines of N\,V, O\,I, Si\,IV, and C\,IV. 
The blue-shifted absorption features in these profiles are consistent with the presence of a high-speed outflow ($>500$\,\kms) along the line-of-sight ($los$) to a hot continuum source.
The lack of absorption features in 
semi-forbidden lines such as SiIII]$\lambda$\,1892.03 and CIII]$\lambda$\,1908.73 is probably due to these being excited in a low-density region that is much larger than the continuum source. 

But the large values of the red-shifts of the 
emission features ($V_{red}(e)$)\footnote{as a fraction of the outflow velocity}, are discrepant from our expectation for a classical P-Cygni profile that results from an outflow 
surrounding a continuum source -- in these profiles, the emission feature is generally centered near or at the systemic velocity (Castor 1970). However, if the outflow velocity 
distribution covers values over a range from $\sim0$ to $V_{max}$, and there is 
substantial line-broadening, e.g. due to microturbulence, the absorption feature may be wide enough to extend to red-shifted velocities thereby reducing the emission-feature's 
blue-wing intensity and thus shifting its centroid substantially redwards of the systemic velocity. But, as can be seen in the predicted spectra from detailed P-Cygni models by van Loon et 
al. (2001: vLKH01), it is only when the absorption feature is extremely saturated, and the microturbulence is a substantial fraction of the maximum outflow 
velocity ($V_{max}$), that one gets values of $V_{red}(e)$ comparable to the blue-shift of the absorption feature, which is located at or near $-V_{max}$  (e.g., see Fig. 9a of vLKH01, 
where $\tau=300$ and the line-broadening parameter $\sigma_v=0.45$)\footnote{In these and other plots of model spectra in vLKH01, all velocities are given as a fraction of $V_{max}$, 
which is set to 1}. When the lines are less optically thick, $V_{red}(e)$ is about $0.3-0.5$ for $\tau=10-100$ (e.g., see Fig. 6a,b of vLKH01, where $\sigma_v=0.2$). 
Since the observed absorption features in our spectra are clearly not heavily saturated, we conclude that a classical P-Cygni model can probably not explain Y\,Gem's UV lines. 

Another model to explain P-cygni profiles is one that was proposed for the PPN, Hen\,3--1475 (S{\'a}nchez Contreras \& Sahai 2001), to explain its H$\alpha$ profile (observed with HST/STIS), 
that also shows blue (red) -shifted absorption (emission) features. This model features a fast, neutral, collimated outflow within more slowly expanding bipolar lobes with dense, 
dusty walls. H$\alpha$ photons produced from a central source pass through gas in the collimated outflow producing the absorption feature, and then are scattered from the dusty walls 
along the $los$, producing a red-shifted emission feature. However, unlike Hen\,3--1475, there is no measurable dust excess in Y Gem (Setal11), so the presence of dusty 
lobes in it is somewhat implausible.

We propose below a new geometric model to explain the large values of $V_{red}(e)$.


\section{Discussion}\label{discus} The short-term stochastic fluctuations in the UV continuum of Y\,Gem appear to be 
similar to the photometric variations seen in other well known classes of accreting binaries, commonly labelled as ``flickering" -- e.g., cataclysmic variables and recurrent novae 
(e.g., Bruch 2015), X-ray binaries with neutron stars and black holes (e.g., van der Klis 2006), symbiotic stars (e.g., Zamanov et al. 2017), and active galactic nuclei (e.g., Pronik et al. 1999). The time-scale of flickering depends on the dimensions of the region where the emission arises. 

Using the light-travel-time corresponding to $\lesssim$20\,s, characteristic of the variations that we have found in Y\,Gem, we estimate that the continuum emitting region is less than 
$\sim0.05$\,au ($8.6$\,\rsun) in size.
Our previous X-ray and UV observations of Y\,Gem also show variability, including both a quasi-periodic and a stochastic component (Setal15). The periodic component seen most clearly in the X-ray data has $P\sim1.3$\,hr, which Setal15 interpreted as being associated with emission from the inner radius of an accretion disk around a sub-solar mass 
($\lsim0.35$\,\ms) main-sequence (MS) companion, although a cold (\teff$<30,000$\,K) white-dwarf (WD) could not be excluded\footnote{A hot WD (as in symbiotic stars) is ruled out by Setal15 because no optical forbidden line emission -- characteristic of symbiotic star spectra -- is seen in Y\,Gem's M8 optical spectrum.}.

However, the bulk outflow speed implied by the velocity offset (from the systemic velocity) of the absorption features  ($\sim500$\,\kms), strongly argues for a main-sequence companion. On both theoretical and empirical grounds, the speed
of an outflow driven from an accretion disk is expected to be of the
order of the escape velocity close to the central accreting
object (Livio 1997), $V_{esc}=620\,\kms\,(M_c/R_c)^{0.5}$, where $M_c$ and $R_c$ are the companion mass and radius in solar units.
Thus the observed bulk expansion velocity of the high-speed outflow is consistent with a solar or sub-solar
MS companion -- e.g., if $M_c\sim0.35$\,\ms~(thus $R_c=0.44$\,\rs: inferred using Table 15.8 in Cox 2000), which is the largest companion mass allowed by the 1.3\,hr orbital 
period seen in Y\,Gem's X-ray light curve (Setal15), then $V_{esc}=550$\,\kms. 
For a cool WD, with a typical mass of 0.6\,\ms, and radius of 0.01\,\rsun, $V_{esc}=4800\,\kms\,(M_c/0.6)^{0.5}\,(0.01/R_c)^{0.5}$, is much larger than the observed outflow velocities. 
Any correction to the outflow velocity due to projection effects is likely to be small, given our geometrical model for the continuum source and outflow (described later in this section). 

Neither of the two blackbodies discussed in \S\,\ref{contmod} fit the properties of a viable stellar companion to the primary. The hotter blackbody is overluminous compared to the expected value 
for a cool WD ($\teff\sim36500$\,K, $L<0.5$\,\ls: Fig. 8 of Miller Bertolami 2016). The cooler blackbody is underluminous for a 
main-sequence star (\teff$\sim9400$\,K, i.e., spectral type early-A, $L\sim30-80$\,\ls). Note of course, that neither of these blackbodies are consistent 
with the temperature and luminosity of a 0.35\,\ms~main-sequence companion. We conclude that both the hot and cool UV components arise in the accretion disk. 

Assuming that the combined luminosity of the hot and cool blackbodies results from accretion, i.e., $L_{acc}=13$\,\ls~during exposures 40--50, then taking 
$L_{acc} \lsim G \mloss_{acc} M_c / R_c$, where $\mloss_{acc}$ is the accretion rate, 
and $M_c\sim0.35$\,\ms~and $R_c=0.44$\,\rs~(as above), we find that 
$\mloss_{acc} > 5\times10^{-7}$\,\my. This relatively large accretion rate makes wind-accretion mechanisms such as Bondi-Hoyle or wind Roche-lobe overflow (e.g., Chen et al. 2017, Huarte-Espinosa et al. 2013) unfeasible 
because Y\,Gem does not have a detectable wind -- the very weak, narrow CO J=2--1 emission line detected towards it likely arises in an extended disk (Setal11); and the infrared-excess 
is negligible or very low (McDonald et al. 2012) implying a mass-loss rate of $\lsim10^{-7}$\,\my~(Snaid et al. 2018). Therefore, either 
the primary overflows its Roche lobe and transfers material to
the accreting star, or accretion occurs within a common-envelope (CE) configuration. For these two accretion modes,  
the binary separation must be small enough and comparable to the primary star's radius -- 1.5\,au for Y\,Gem's luminosity and effective temperature of 5800\,\ls~and 2800\,K, respectively (Setal11). However, since we observe the accretion luminosity in the UV, a 
CE scenario -- in which the radiation would be trapped within the envelope -- is less likely. We conclude that Roche lobe overflow 
is the most likely accretion mode for Y\,Gem.
 

The absence of molecular H$_2$ lines (e.g., prominently detected in accreting T--Tauri stars: Ingleby et al. 2011) in Y\,Gem's UV spectrum is striking, especially since the detection of the 6.4\,keV FeI line in its X-ray spectrum implies the presence of a neutral disk in it. 
The two main mechanisms for exciting  H$_2$ lines are fluorescence due to \lya~pumping (e.g. Yang et al. 2011) or collisional excitation by hot electrons (e.g. Ingleby et al. 2009).  If the disk in Y\,Gem is composed primarily of molecular gas, then the absence of the H$_2$ lines suggests that there are large temporal variations in the total amount of neutral material in the disk.  Alternatively, H$_2$ line emission could be present but blocked from the observer's view by an optically-thick continuum emitting region of hot gas.
Otherwise, the disk is primarily composed of atomic gas.

\begin{figure*}[hbt]
\resizebox{0.5\textwidth}{!}{\includegraphics{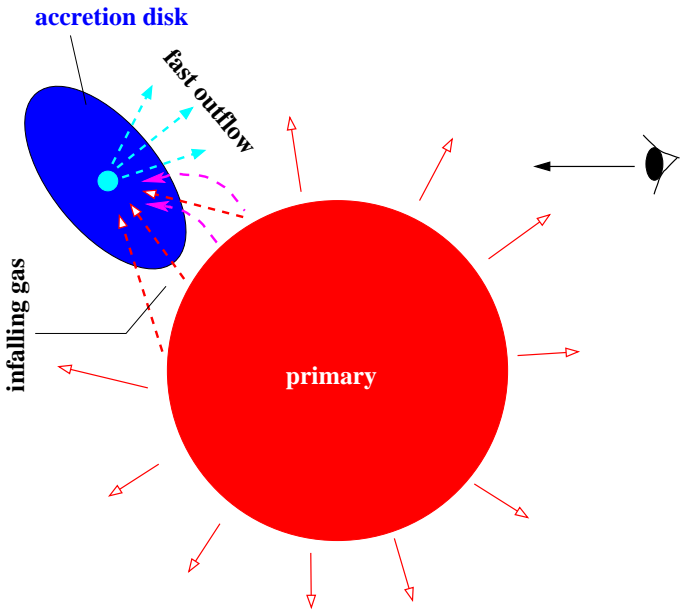}}
\caption{Schematic geometry of a model (not to scale) to explain the Doppler shifts of the absorption and emission features observed in the UV spectra of Y\,Gem. A hot accretion disk (blue ellipse) around a compact companion (cyan circle) captures part of the material (red arrows) in an outflow or Roche-lobe overflow from the primary AGB star (red circle), via two possible infall streams (dashed red/pink arrows) -- these produce red-shifted emission features. The accretion disk powers a fast outflow (cyan arrows) that absorbs UV photons from the disk, producing blue-shifted absorption features.
}
\label{schematic}
\end{figure*}

We propose a simple geometric  
model that can explain the Doppler-shifts of the absorption and emission features (Fig.\,\ref{schematic}). In this model, material from the primary AGB star (either from an outflow or Roche-lobe overflow) is gravitationally focussed towards the companion, producing a hot accretion disk. The emission features arise in the infalling material which gets heated either from internal shocks (e.g., due to density and velocity variations within the stream) and/or friction due to passage through the accretion disk's ``atmosphere". The accretion disk produces an outflow that is seen in absorption against the continuum emission from the accretion disk. 
Sufficiently hot regions in the disk may also produce UV emission lines which would be centered roughly around the systemic velocity.  Since we do not see a signature of such emission in our spectra, the disk emission lines must be relatively faint compared to those from the outflow, and the size of the hot disk region must be small compared with that of the outflow.  

The large observed values of the blue- and red- Doppler shifts of the absorption and emission features, respectively, are naturally explained in this model because both features are associated with material moving in the 
gravitational well of the central star near the accretion disk's inner regions. 
The emission features from high-excitation lines (C\,IV, Si\,IV) are generated closer to the disk, in the infall's hottest parts, whereas the low-excitation Mg\,II lines are generated in a region further away, hence the red-shift of the Mg\,II lines is much lower than that of the high-excitation lines.

The correlation between the variations seen in the FUV continuum and Si\,IV absorption line features suggests that both of these are associated with variations in the accretion rate.
The mismatch between the epoch during which the Si\,IV(1) absorption line equivalent-width peaks, and those at which $L(h)$ peaks (in exposure 50), 
confirms that these arise in spatially-separated structures. In this model, the long-term UV and X-ray variations noted by Setal15 may result from orbital motion as varying segments of the disk are eclipsed by the AGB star. 

\begin{figure*}[hbt]
\rotatebox{270}{\resizebox{0.5\textwidth}{!}{\includegraphics{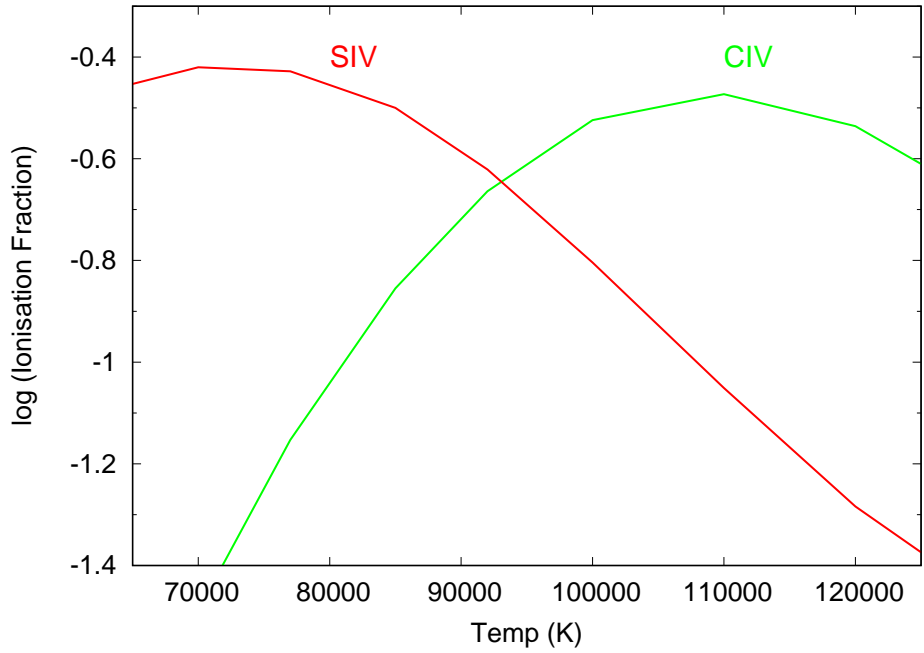}}}
\caption{The ionization fractions of C\,IV ({\it green}) and Si\,IV ({\it red}) for a range of temperatures, for a cloud of density $10^3$\,cm$^{-3}$.
}
\label{ionis_frac}
\end{figure*}

We also find an anti-correlation between the fluxes of the C\,IV and Si\,IV lines -- e.g., over the period covered by exposures 10 to 60, the Si\,IV flux shows a general increase, 
whereas the C\,IV flux shows 
a general decrease (Fig.\,\ref{contlumtemp}b). A plausible explanation for this anti-correlation is that it results from  
changes in the ionization fractions of each of these species as a function of temperature. Our simple CLOUDY (Ferland et al. 2013) modeling using uniform low-density spherical plasma clouds, with a wide range of plausible densities, $\sim10-10^5$\,cm$^{-3}$, show that the ionization fractions of C\,IV and Si\,IV are anti-correlated for temperatures in the range $(0.7-1.1)\times10^5$\,K  -- i.e., a decrease in temperature results in an increase (decrease) in the fractional population of the Si\,IV (C\,IV) ionization state (Fig.\,\ref{ionis_frac}). 
We note that within this temperature range, the Si\,IV and C\,IV ionization states are also well-populated -- they represent the second most-populated states: their populations  
differ, at e.g., $T\sim0.9\times10^5$\,K, by factors 2.4 and 3.3, respectively, from that of the most populated ones.

The anti-correlation between the blue-shift for the Si\,IV(1) absorption feature and the red-shift of the Si\,IV(1) emission feature may be explained as follows. In our model, 
the P-Cygni feature is due to the sum of the features produced by (i) absorption of continuum by the outflow, and (ii) emission from the infalling stream (which sits on top of the underlying continuum). When the outflow velocity increases, the absorption feature shifts bluewards as a whole, thus making the blue-side of the emission feature stronger by decreasing the underlying absorption -- the net result is that the observed red-shift of the emission feature becomes smaller.

In order to explain the observed short-term variations in the line-widths, red- and blue-shifts, and fluxes, we propose that the infall and outflow streams experience short-term temporal variations  in the densities, temperatures, velocities and the velocity gradients in these streams, that are likely stochastic. The relatively large observed line-widths -- e.g., $\sim335-515$\,\kms ($\sim775-1000$\,\kms) for the emission features in 
Si\,IV(1) (Ly$\alpha$), depending on the epoch -- are likely a consequence of velocity gradients (due to both turbulent and systematic motion) and 
the presence of a range of inclinations, relative to the $los$, within the outflow and infall streams. 


\section{CONCLUSIONS}\label{conclude}

Using the {\it Hubble Space Telescope}, we have carried out UV spectroscopic observations of the late-M star, Y\,Gem -- the most prominent member of a 
class of AGB stars that are sources of strong and variable UV and X-ray emission, likely resulting from 
accretion activity due to the presence of a binary companion. 

\begin{enumerate}
\item Y\,Gem shows the presence of strong emission in the FUV ($\sim$1120\,\AA--1700\,\AA) and NUV ($\sim$1600\,\AA--3150\,\AA) bands, both in the continuum 
as well as in lines such as Ly$\alpha$, C\,IV\,$\lambda\lambda$1548,1551, Si\,IV\,$\lambda\lambda$1394,1403, and 
Mg\,II\,$\lambda\lambda2796,2803$. 
\item The UV continuum shows short-term stochastic time variations (on time-scales of  $\lesssim20$\,s) -- this flickering phenomenon is characteristic of the presence of 
an active accretion disk. We also find a long-term trend (overall decline) in the FUV continuum over a period of $\sim1.7$\,hr.
\item The continuum can be modeled as a sum of a hotter and a cooler blackbody component. The luminosities of these components, as derived from the best-modeled pair of near-contemporaneous 
FUV and NUV spectra, are $\sim6.8$\,\ls~and $6.3$\,\ls, and the temperatures are $\sim3.7\times10^4$\,K and $\sim10^4$\,K, respectively. The changes in the FUV continuum 
appear to be related to similar changes in the luminosity of the hotter blackbody; the temperature is significantly less variable.
\item Neither of the two blackbodies fit the properties of a viable stellar companion (white-dwarf or main-sequence) to the primary, suggesting that both the hot and cool UV components arise in the accretion disk.
\item Lines from species such as Si\,IV and C\,IV show prominent P-Cygni-type profiles, with emission and absorption features that are red- and blue- shifted by  
$\sim$500\,\kms~from the systemic velocity. The relatively large red-shifts of the emission features are not consistent with a classical P-Cygni profile that results from an outflow 
surrounding a continuum source.
\item Our model for these (and previous) observations is that material from the primary star 
is gravitationally captured by a companion, producing a hot accretion disk. The latter powers a fast outflow 
that produces blue-shifted features due to absorption of UV continuum emited by the disk, whereas the red-shifted emission features arise in heated infalling material from the primary.
\item The outflow velocities support a previous inference that Y\,Gem's companion is a low-mass main-sequence star.
\item The combined luminosity of the hot and cool blackbodies require a relatively large accretion rate, $> 5\times10^{-7}$\,\my; since 
the primary in Y\,Gem's has negligible (or very weak) mass-loss, wind-accretion modes are unfeasible, 
and between common-envelope and Roche lobe overflow accretion modes, the latter is preferred.
\end{enumerate}


\acknowledgements
We acknowledge helpful discussions with Eric Blackman related to binary accretion modes. 
RS's contribution to the
research described here was carried out at the Jet Propulsion Laboratory, California Institute of Technology, under a
contract with NASA.  CSC has been partially supported by the Spanish MINECO through grant AYA2016-75066-C2-1-P and by 
the European Research Council through ERC grant 610256: NANOCOSMOS. JSF acknowledges support
from the Spanish MINECO through grant AYA2014-54348-C3-2-R. The National Radio Astronomy Observatory is a facility of the National Science Foundation 
operated under cooperative agreement by Associated Universities, Inc.

\clearpage

\begin{table*}
\vskip -0.1in
\caption{Log of Observations}
\vspace{-0.1in}
\begin{tabular}{lcclllll}
\hline
Dataset   & Exp\,\#\footnotemark[1] & Grating & Start Time & Exp.Time\footnotemark[2] \\
OD9C01010 & 10 & G140L & 2016-10-11 21:56:34 & 347.020 \\
OD9C01020 & 20 & G140L & 2016-10-11 22:05:47 & 347.020 \\
OD9C01030 & 30 & G230L & 2016-10-11 23:02:57 & 312.126 \\
OD9C01040 & 40 & G230L & 2016-10-11 23:12:10 & 304.769 \\
OD9C01050 & 50 & G140L & 2016-10-11 23:26:06 & 517.020 \\
OD9C01060 & 60 & G140L & 2016-10-11 23:38:09 & 517.020 \\
OD9C01070 & 70 & G230L & 2016-10-12 00:38:23 & 503.600 \\
OD9C01080 & 80 & G230L & 2016-10-12 00:50:26 & 486.315 \\
\hline
\end{tabular}
\footnotetext[1]{Exposure No.}
\footnotetext[2]{Exposure Time (s).}
\label{logobs}
\end{table*}

\begin{table*}
\vskip -0.1in
\caption{Properties of Selected Observed Lines from Exposures 20 (G140L) and 30 (G230L)}
\vspace{-0.1in}
\begin{tabular}{lclllll}
\hline
Name   & Rest $\lambda$ & Emiss./Abs.\footnotemark[1]& Obs. $\lambda$ & $V_{hel}$ & FWHM\footnotemark[2] &  Flux  \\
       & \AA            &                            & \AA            & $\kms$    & $\kms$               &  $10^{-13}\,\fluxu$  \\
\hline
Ly $\alpha$                & 1215.670 & Emiss. & 1218.573 $\pm$ 0.008  &  716  $\pm$ 2  &  774 $\pm$ 4  &   54.5 $\pm$ 0.4 \\
Si\,IV(1)\footnotemark[3]  & 1393.756 & Abs.   & 1391.200 $\pm$ 0.030  &  $-$549 $\pm$ 7  &  565 $\pm$ 17 &  $-$4.70 $\pm$ 0.15 \\
Si\,IV(1)\footnotemark[3]  & 1393.756 & Emiss. & 1395.657 $\pm$ 0.023  &  409  $\pm$ 5  &  436 $\pm$ 11 &   6.84 $\pm$ 0.19 \\      
Si\,IV(2)\footnotemark[4]  & 1402.770 & Abs.   & 1400.366 $\pm$ 0.037  &  $-$514 $\pm$ 8  &  322 $\pm$ 19 &  $-$2.43 $\pm$ 0.13 \\
Si\,IV(2)\footnotemark[4]  & 1402.770 & Emiss. & 1404.613 $\pm$ 0.030  &  394  $\pm$ 6  &  518 $\pm$ 16 &   6.73 $\pm$ 0.23 \\
C\,IV(1,2)\footnotemark[5] & 1549.490 & Emiss. & 1553.249 $\pm$ 0.026  &  727  $\pm$ 5 &  941 $\pm$ 11 &   27.7 $\pm$ 0.5 \\
C\,IV(1,2)\footnotemark[5] & 1549.490 & Abs.   & 1545.309 $\pm$ 0.060  &  $-$809 $\pm$ 12 &  588 $\pm$ 30 &  $-$3.42 $\pm$ 0.21 \\
Si\,III]                   & 1892.030 & Emiss. & 1893.453 $\pm$ 0.026  &  225  $\pm$ 4  & 915  $\pm$ 9  &   34.5 $\pm$ 0.5 \\
C\,III]\footnotemark[6]    & 1908.734 & Emiss. & 1910.320 $\pm$ 0.038  &  249  $\pm$ 6  & 1022 $\pm$ 14 &   23.4 $\pm$ 0.4 \\
Mg\,II(1)\footnotemark[3]  & 2796.352 & Emiss. & 2798.318 $\pm$ 0.013  &  211  $\pm$ 1  &  537 $\pm$ 3  &   49.8 $\pm$ 0.4 \\
Mg\,II(2)\footnotemark[4]  & 2803.531 & Emiss. & 2804.983 $\pm$ 0.033  &  155  $\pm$ 4  &  511 $\pm$ 7  &   27.8 $\pm$ 0.4 \\
\hline
\end{tabular}
\footnotetext[1]{Emiss.= Emission, Abs.=Absorption}
\footnotetext[2]{Corrected for instrumental resolution}
\footnotetext[3]{1st line in a resolved (or partially resolved)  doublet}
\footnotetext[4]{2nd line in a resolved (or partially resolved)  doublet}
\footnotetext[5]{mean wavelength of unresolved C\,IV doublet with $\lambda_{0}$=1548.20\,\AA~and 1550.78\,\AA}
\footnotetext[6]{unresolved blend with [C\,III] transition at $\lambda_{0}$=1906.683\,\AA}
\label{lines}
\end{table*}
\clearpage

%

\end{document}